# SYNDROME-TRELLIS DECRYPTION IN CODE-BASED POST-QUANTUM CRYPTOGRAPHY

Meir Ariel

School of Electrical and Computer Engineering,
Faculty of Engineering, Tel Aviv University, Tel Aviv, Israel

***ABSTRACT***

*We present a novel post-quantum cryptographic framework based on masked high-memory convolutional codes in the Niederreiter paradigm. The public key is derived from a high-memory polynomial parity-check matrix augmented by dense random masking and additional invertible linear transformations. The resulting matrix exhibits strong random-like properties, effectively concealing its algebraic structure and resisting known structural and algebraic attacks. Legitimate recipients retain polynomial-time decoding capability, whereas adversaries appear to be limited to generic information-set decoding with exponential complexity. Under standard cryptanalytic estimates, the proposed construction achieves security margins exceeding those of classical McEliece- and Niederreiter-based systems by factors greater than $2^{200}$. Beyond its enhanced security profile, the construction offers considerable design flexibility, supporting arbitrary plaintext lengths, linear-time decryption, and a uniform per-bit computational cost, enabling efficient scalability to very long messages. Practical implementation is facilitated by a parallel array of syndrome-trellis decoders, dynamically instantiated according to candidate syndrome values, enabling efficient hardware and software implementations with high throughput. Overall, the proposed scheme is a promising candidate for robust, scalable, quantum-resistant public-key cryptography.*

## 1. INTRODUCTION

Code-based cryptography, first proposed by Robert McEliece in 1978 [1], employs binary Goppa codes to construct a public-key encryption scheme. Its central idea is to disguise an error-correcting code through invertible linear transformations, thereby producing a public key that conceals the underlying algebraic structure. During encryption, the sender deliberately introduces controlled errors into the codeword, making decryption computationally infeasible for adversaries, while the legitimate recipient—who possesses the private key—can efficiently correct the errors and recover the original plaintext.

Since its inception, numerous variants of the McEliece cryptosystem have been proposed, typically by replacing Goppa codes with alternative code families [2–3]. However, most of these alternatives have failed to preserve the strong security guarantees of the original construction.

Consequently, Classic McEliece advanced to Round 4 of the NIST Post-Quantum Cryptography (PQC) standardization process [4], although it was ultimately not selected for standardization. Comprehensive surveys of code-based cryptography and its role in the PQC landscape are provided in [5–6].

Despite its robustness and long-standing security record, Classic McEliece suffers from several practical limitations. Its reliance on fixed Goppa codes imposes relatively rigid error-correction capabilities, limits adaptability across different security levels, and complicates deployment in diverse application environments. To address these limitations, Ariel [7–8] proposed a novel post-quantum cryptographic framework based on noise-enhanced high-memory convolutional codes. This approach enables dynamic code construction and selection while enhancing public-key security through dense, random-like generator matrices that conceal structural information more effectively than low-density or structurally recognizable matrices [9]. Moreover, the use of powerful convolutional codes with greater error-correcting capability permits the deliberate introduction of higher noise levels, thereby increasing resistance to cryptanalytic attacks by factors exceeding $2^{200}$ (depending on the selected parameters) compared with Classic McEliece. In addition, the framework supports scalability to very long plaintexts, linear-time decoding, and efficient hardware and software implementations with uniform per-bit computational complexity. The "Niederreiter Cryptosystem" was first proposed by Harald Niederreiter in 1986 [10] as a variant of the McEliece encryption scheme, using the parity-check matrix of a linear error-correcting code as the public key rather than the generator matrix. In this formulation, the plaintext is a *coset leader* $\boldsymbol{e}$, i.e., the minimum-Hamming-weight error vector associated with a given coset. The ciphertext is the syndrome $\boldsymbol{s}$, obtained by multiplying $\boldsymbol{e}$ by the public parity-check matrix. Unlike the McEliece system, the ciphertext contains no additional injected errors. Nevertheless, the scheme remains probabilistic because the plaintext itself is a randomly selected error vector, while the public parity-check matrix appears random and has no known efficient decoder.

The decryption problem in the Niederreiter cryptosystem is therefore equivalent to the classical syndrome decoding problem, which is known to be NP-hard. Moreover, as noted in [11], McEliece and Niederreiter cryptosystems are considered equivalent from the security perspective when corresponding parameter sets are used. Consequently, the Niederreiter scheme inherits many of the same limitations as Classic McEliece, including reliance on Goppa codes with rigid parameter selection, relatively limited error correction capabilities, and algebraic decoding procedures that are not well suited for efficient hardware implementations.

In this paper we extend the framework presented in [7–8] by employing masked high-memory parity-check matrices derived from high-performance convolutional codes as public keys, thereby overcoming several limitations of the Niederreiter approach. The proposed post-quantum cryptographic framework offers the following advantages:

**Diverse Code Selection**: A broad family of convolutional codes can serve as public keys, enabling flexible adaptation to different performance and security levels.

**Enhanced Public Key Security**: The proposed scheme generates dense, random-like parity-check matrices that effectively conceal structural information. Moreover, the use of powerful convolutional codes with high error-correcting capacity allows the selection of error vectors (plaintexts) with significantly larger Hamming weights, thereby increasing resistance to cryptanalytic attacks by factors exceeding $2^{200}$ (depending on the selected parameters) compared with the classical Niederreiter scheme.

**Scalability and Linear-Time Decoding**: Unlike fixed-dimension block code constructions, the proposed framework supports plaintexts of arbitrary length. Decoding complexity grows linearly with plaintext size, enabling scalability without loss of efficiency.

**Efficient Implementation**: Decryption is performed by an array of unique directed-graph decoders, referred to as *Syndrome Trellises*, that identify the correct plaintext while supporting efficient and straightforward hardware and software implementations.

The remainder of this paper is organized as follows. Section 2 describes the construction of the public and private keys based on masked high-memory convolutional codes, which form the foundation of the proposed framework. Section 3 presents the encryption process, based on the selection of random error vectors, which is central to both security and reliable decryption. Section 4 introduces the construction of an array of syndrome-trellis decoders, and Section 5 details the corresponding decryption procedure performed by the legitimate recipient. Section 6 provides a worked implementation example, while Section 7 analyses potential attacks and evaluates the cryptanalytic resistance and computational complexity of the proposed scheme. Finally, Section 8 concludes the paper and summarizes the main findings.
The principal contributions of this paper are as follows:

- Extension of the previously proposed masked high-memory convolutional-code cryptographic framework from generator matrices to parity-check matrices, thereby enabling a Niederreiter-type post-quantum cryptosystem.
- A novel directed-graph decryption algorithm for masked parity-check cryptosystems, based on high-memory convolutional codes.
- A practical method for constructing syndrome trellises corresponding to arbitrary received syndromes, enabling efficient maximum-likelihood decoding in polynomial time.
- A polynomial-time ambiguity-resolution technique that identifies the correct plaintext by minimum-weight decoding over multiple candidate syndrome trellises.
- A complexity and security analysis showing that legitimate decryption scales linearly with the plaintext length while the best-known attacks remain exponential.
- A complete worked implementation example illustrating key generation, encryption, syndrome-trellis construction, and decryption.

## 2. PUBLIC AND PRIVATE KEY CONSTRUCTION

Throughout this paper, the notation follows that of [8] and is also summarized here for completeness. Binary vectors are denoted by bold lowercase letters (e.g., $\boldsymbol{a}$, $\boldsymbol{b}$) while their corresponding polynomial representations are written as $\boldsymbol{a}(x)$, $\boldsymbol{b}(x)$. The notations $\boldsymbol{a}$ and $\boldsymbol{a}(x)$ are used interchangeably whenever context is clear. The Hamming weight of a vector $\boldsymbol{a}$ is denoted by wt($\boldsymbol{a}$).Standard coding-theorynotation is adopted throughout: $\boldsymbol{e}$ denotes an error vector,$\boldsymbol{s}$ denotes a syndrome, and polynomial generator or parity-check matrices associated with convolutional codes are denoted by uppercase letters such as $G(x)$and $H(x)$, with their corresponding scalar representations written as $G$ and $H$. Calligraphic letters, such as $\mathcal{D}$ and $\mathcal{L}$, denote sets of binary vectors.

To establish a clear foundation for the proposed cryptosystem, we first briefly review the classical Niederreiter cryptosystem:

**NiederreiterKey Generation (Bob)**

Let $C$ be a binary $(n, k)$ linear code capable of correcting up to $t$ errors, with parity-check matrix

$$H \in \mathbb{F}_2^{(n-k)\times n}$$

Bob randomly selects:

- a nonsingular matrix $S \in \mathbb{F}_2^{(n-k)\times(n-k)}$,
- a permutation matrix $P \in \mathbb{F}_2^{n\times n}$.

He then publishes the obfuscated parity-check matrix:

$$\widehat{H} = SHP \tag{1}$$

The public key is $(\widehat{H}, t)$, while the private key is $(S, H, P)$.

**Niederreiter Encryption (Alice)**

To encrypt a message Alice selects an error vector $\boldsymbol{e} \in \mathbb{F}_2^n$ with Hamming weight wt($\boldsymbol{e}$) $\leq t$, and computes the syndrome:

$$\boldsymbol{s} = \widehat{H}\boldsymbol{e}^T \tag{2}$$

The ciphertext is $\boldsymbol{s} \in \mathbb{F}_2^{n-k}$.

**Niederreiter Security (Eve)**

Given $\widehat{H}$ and the intercepted $\boldsymbol{s}$, Eve must solve

$$\boldsymbol{s} = \widehat{H}\hat{\boldsymbol{e}}^T \tag{3}$$

for a low-weight vector $\hat{\boldsymbol{e}}$. This constitutes an instance of the classical syndrome decoding problem, which is believed to be computationally intractable for suitably chosen parameters. Since $\widehat{H}$ appears random and admits no efficient decoding algorithm, recovering $\hat{\boldsymbol{e}}$ is computationally infeasible.

**Niederreiter Decryption (Bob)**

Bob first computes

$$\boldsymbol{s}' = S^{-1}\boldsymbol{s} = HP\boldsymbol{e}^T \tag{4}$$

Let $\boldsymbol{e}' = \boldsymbol{e}P$, then

$$\boldsymbol{s}' = H\boldsymbol{e}'^T \tag{5}$$

Using the decoder associated with the code $C$, Bob recovers the unique low-weight vector $\boldsymbol{e}'$, and finally computes

$$\hat{\boldsymbol{e}} = \boldsymbol{e}'P^{-1} \tag{6}$$

Correctness follows directly from the decoding capability of $C$ for all error vectors satisfying wt($\boldsymbol{e}$) $\leq t$. Security relies on selecting a code such that the public matrix $\widehat{H}$ does not revealexploitable algebraic structure.

We now describe the construction of the public and private keys in the proposed framework, based on a masked high-memory parity-check matrix derived from a convolutional code, which forms the foundation of the proposed cryptographic scheme.

Denote by $\boldsymbol{p}_i(x)$ a binary polynomial of memory at most$p$

$$\boldsymbol{p}_i(x)= \sum_{j=0}^{p} a_j x^j, \quad a_j \in \mathbb{F}_2 \tag{7}$$

A Convolutional Code (CC) is defined as a set of $n$ constituent polynomials, commonly structured into the form of a *polynomial generator matrix*

$$G(x) = [\boldsymbol{p}_0(x), \boldsymbol{p}_1(x), \ldots, \boldsymbol{p}_{n-1}(x)] \tag{8}$$

The matrix $G(x)$ determines both the code rate and the error-correction capability of the CC, typically characterized by its free distance $d_{free}$.

The polynomial matrix $G(x)$ admits an equivalent infinite-dimensional scalar representation $G$, whose coefficient $k \times n$ matrices $\boldsymbol{g}_i$ corresponds to the coefficients of $x^i$:

$$G = \begin{bmatrix} \boldsymbol{g}_0 & \boldsymbol{g}_1 & \boldsymbol{g}_2 & \cdots & \boldsymbol{g}_p & \mathbf{0} & \mathbf{0} & \mathbf{0} & \cdots \\ \mathbf{0} & \boldsymbol{g}_0 & \boldsymbol{g}_1 & \cdots & \boldsymbol{g}_{p-1} & \boldsymbol{g}_p & \mathbf{0} & \mathbf{0} & \cdots \\ \mathbf{0} & \mathbf{0} & \boldsymbol{g}_0 & \cdots & \boldsymbol{g}_{p-2} & \boldsymbol{g}_{p-1} & \boldsymbol{g}_p & \mathbf{0} & \cdots \\ \vdots & & & & & & & & \end{bmatrix} \tag{9}$$

For example, for $G(x) = [1+x+x^2, 1+x^2]$, the corresponding coefficient vectors are

$$\boldsymbol{g}_0 = [1\ 1], \boldsymbol{g}_1 = [10], \boldsymbol{g}_2 = [1\ 1].$$

A CC mayalso be defined by its $(n-k) \times n$polynomial parity-check matrix$H(x)$ satisfying

$$G(x)\, H(x)^{\mathrm{T}} = \mathbf{0} \tag{10}$$

For rate-½ CCs, $H(x)$ is obtained simply by reversing the order of the generator polynomials. Like $G(x)$, $H(x)$ admits a scalar representation $H$, whose coefficients$(n-k) \times n$ matrices $\boldsymbol{h}_i$correspond to the coefficients of $x^i$.

$$H = \begin{bmatrix} \boldsymbol{h}_0 & \mathbf{0} & \mathbf{0} & \cdots \\ \boldsymbol{h}_1 & \boldsymbol{h}_0 & \mathbf{0} & \\ \boldsymbol{h}_2 & \boldsymbol{h}_1 & \boldsymbol{h}_0 & \\ \vdots & \boldsymbol{h}_2 & \boldsymbol{h}_1 & \\ \boldsymbol{h}_{p-1} & \vdots & \boldsymbol{h}_2 & \\ \boldsymbol{h}_p & \boldsymbol{h}_{p-1} & \vdots & \\ \mathbf{0} & \boldsymbol{h}_p & \boldsymbol{h}_{p-1} & \\ \mathbf{0} & \mathbf{0} & \boldsymbol{h}_p & \\ \vdots & & & \end{bmatrix} \tag{11}$$

For simplicity, we focus on CCs of rate $1/n$ (i.e., $n$–$k$=1). The proposed construction, however, is readily applicable to arbitrary CCs, including punctured codes.

Each constituent polynomial $\boldsymbol{p}_i(x)$is multiplied by a *high-memory polynomial*$\boldsymbol{q}(x)$of degree $q$,

$$\boldsymbol{q}(x) = \sum_{j=0}^{q} b_j x^j, b_j \in \mathbb{F}_2 \tag{12}$$

yieldinga high-memory parity-check matrix

$$H'(x) = H(x)\boldsymbol{q}(x) = [\boldsymbol{p}_0(x)\boldsymbol{q}(x), \boldsymbol{p}_1(x)\boldsymbol{q}(x), \ldots, \boldsymbol{p}_{n-1}(x)\boldsymbol{q}(x)] \tag{13}$$

The proposed framework imposes no strict constraints on the choiceof $\boldsymbol{q}(x)$. In practice, however, the parameters $p$ and $q$ are chosen such that

$$p + q > 200, p \ll q$$

thereby providing a very large effective memory. The corresponding scalar representation$H'$therefore exhibits substantially longer column run lengths and significantly higher density than the original parity-check matrix $H$

$$H' = \begin{bmatrix} \boldsymbol{h}_0 & \mathbf{0} & \mathbf{0} & \cdots \\ \boldsymbol{h}_1 & \boldsymbol{h}_0 & \mathbf{0} & \\ \boldsymbol{h}_2 & \boldsymbol{h}_1 & \boldsymbol{h}_0 & \\ \vdots & \boldsymbol{h}_2 & \boldsymbol{h}_1 & \\ \boldsymbol{h}_{p+q-1} & \vdots & \boldsymbol{h}_2 & \\ \boldsymbol{h}_{p+q} & \boldsymbol{h}_{p+q-1} & \vdots & \\ \mathbf{0} & \boldsymbol{h}_{p+q} & \boldsymbol{h}_{p+q-1} & \\ \mathbf{0} & \mathbf{0} & \boldsymbol{h}_{p+q} & \\ \vdots & & & \end{bmatrix} \tag{14}$$

The desired plaintext length $N$ determines the dimensions of a truncated matrix$H'$. Specifically, the number of columns is fixed at$N$, while the number of rows, denoted by$N - K$, depends on both the coderate and the overall memory length of the CC. For rate-$1/n$CCs

$$N - K = N + p + q - \frac{N}{n}$$

Increasing the degree of the high-memory polynomial $\boldsymbol{q}(x)$not only increases the density and apparent randomness of the resulting public matrix, making it more suitable for subsequent concealing transformations, but also enlarges both the associated syndrome and decoder state spaces. Together, these effects increase the complexity of structural analysis and decoding attacks, thereby strengthening the cryptographic security of the proposed framework.

For a CC with memory $m$, a Viterbi decoder employs a trellis with $2^m$states. Appending $m$ zero bits to the encoder input forces the trellis into a unique terminal state. Similarly, in the proposed framework, decoding is formulated as a search for the most likely error vector within a given coset. This search is performed on a directed graph representing the coset, known as a syndrome trellis [12–13]. The last $m$ syndrome bits drive the syndrome trellis to a unique terminal state, as described in Section 4. Since each coset has its own syndrome trellis, the graph is generated dynamically, section by section, according to the syndrome values. Such on-the-fly generation is essential because the number of cosets—and hence syndrome trellises—is astronomically large, making precomputation impractical.

As follows from Equation (14), each group of $n$ columns of $H'$is obtained by shifting the preceding group downward by$n - k$ bits. Although multiplication by the high-memory polynomial $\boldsymbol{q}(x)$significantlyincreases column run lengths of$H'$, permutation alone cannot conceal its structure, since the remaining zero sequences still reveal recognizable diagonal patterns. To eliminate this structure and maximize density, we introduce an $(N-K) \times N$ *masking matrix* $\widetilde{H}$ of rank $l$. Each column of $\widetilde{H}$ is selected randomly from the linear span of a predefined set $\mathcal{L}$ of $l$ random binary vectors of length $N-K$. Denoting this span by

$$\mathrm{LS}(\mathcal{L}) = \{\boldsymbol{l}_0, \boldsymbol{l}_1, \ldots, \boldsymbol{l}_{2^l-1}\} \tag{15}$$

where each $\boldsymbol{l}_i$ is a distinct linear combination of the vectors in $\mathcal{L}$, the masked parity-check matrix is

$$H' \oplus \widetilde{H}$$

where $\oplus$ denotes bitwise addition over $\mathbb{F}_2$. This operation destroys the original diagonal structure of $H'$, producing a dense random-like matrix. Since $H'$ has full rank, $\widetilde{H}$ may be chosen so that $H' \oplus \widetilde{H}$ also has full rank, enabling subsequent invertible transformations. The public key matrix $\widehat{H}$ is then defined as

$$\widehat{H} = S\left(H' \oplus \widetilde{H}\right)P \tag{16}$$

where $S$ is a random nonsingular $(N-K) \times (N-K)$ matrix and $P$ is a random $N \times N$ permutation matrix. The matrix $S$ obscures the correspondence between error vectors (plaintexts) and syndromes (ciphertexts), while $P$ conceals the column positions. Consequently, the transformation from $H$ to $\widehat{H}$ combines reversible operations $\boldsymbol{q}(x)$, $S$ and $P$ with the random masking operation $\widetilde{H}$, yielding a dense random-like public matrix. We refer to $\widehat{H}$ as the High-Memory Masked Parity-Check (MPC) matrix.

To ensure reliable decryption despite the syndrome perturbation introduced by masking, the correct plaintext is identified as the candidate coset leader with minimum Hamming weight, while all other candidates behave essentially as random coset leaders. An error vector $\boldsymbol{e}$ is generated by independently injecting errors into an all-zero sequence of length $N$ with probability $\varepsilon$, yielding

$$\mathrm{wt}(\boldsymbol{e}) \approx \varepsilon N \tag{17}$$

In contrast, a random coset leader $\tilde{\boldsymbol{e}}$ typically has Hamming weight

$$\mathrm{wt}(\tilde{\boldsymbol{e}}) \approx \delta \tag{18}$$

where $\delta$ denotes the typical nearest-neighbor distance given by the Gilbert bound

$$E\left(\frac{\delta}{N}\right) \approx 1 - \frac{K}{N} \tag{19}$$

and $E(\cdot)$ denotes the binary entropy function. Incorrect masking vectors produce essentially random cosets. Consequently, their minimum-weight representatives concentrate around $\delta$. Since

$$\varepsilon N << \delta$$

the probability that an incorrect candidate attains weight close to $\varepsilon N$ decreases exponentially with $N$. The parameter $\varepsilon$ is chosen to balance reliable decryption against cryptanalytic resistance. An important advantage of this approach is that the error vector is statistically random. Correct decryption, however, is not guaranteed, since localized clusters of errors may occasionally cause decoding failure. The probability of such failures can be estimated in advance, allowing $\varepsilon$ to be selected accordingly, as discussed in Section 7. Finally, the decoder described in Section 4 performs optimum decoding and therefore requires no prior knowledge of either the number or distribution of injected errors.

Decryption failure may be eliminated by constraining the error injection process as follows: Partition the error vector into "windows" of length

$$w = n(p+1)$$

For a CC with free distance $d_{\text{free}}$, any two trellis paths that diverge and remerge within such window differ in at least $d_{\text{free}}$ positions.Consequently, up to

$$t_i \le \frac{d_{free} - 1}{2} \tag{20}$$

errors can be corrected reliably within the $i$-thwindow. The resulting error vector therefore has Hamming weight

$$\text{wt}(\boldsymbol{e}) = \sum_{i=1}^{\frac{N}{w}} t_i \tag{21}$$

Under this constraint, correct decoding is guaranteed. Even so, the number of possible plaintexts remains astronomically large. Assuming, for simplicity, a fixed value of$t_i$for all windows. The total number of possible plaintexts is

$$\binom{w}{t_i}^{N/w}$$

For the CC parameters considered in Section 7, namely$w$ = 30, $d_{\text{free}}$=20 (hence 9 correctable errors per window), $t_i$ = 7 and $N$ = 8200

$$\binom{30}{7}^{273} \approx 2 \times 10^{1722}$$

demonstrating the enormous plaintext space obtainable even under constrained error injection. Nevertheless, this approach is less desirable than fully random error generation, since imposing structure on the plaintext may introduce potential avenues for cryptanalytic attacks.

The public and private keys are therefore defined as follows:

**Public key:** $\{\widehat{H}, \varepsilon\}$

**Private key:** $\{S, P, H, \boldsymbol{q}(x), \widetilde{H}\}$

Decryption proceeds in several stages. First, the inverse transformation $S^{-1}$ is applied to the received syndrome. The masking introduced by $\widetilde{H}$must then be removed; however, this operation is not fully reversible. Although $\widetilde{H}$ is known to the legitimate recipient, the plaintext error vector is unknown, and consequently the corresponding linear combination of masking vectors added during encryption is also unknown, resulting in ambiguity.

Next, the obfuscation introduced by the high-memory polynomial $\boldsymbol{q}(x)$is reversed, since practical decoding can be performed only with the original parity-check matrix $H$, not its high-memory counterpart $H'$.Accordingly, all candidate syndromes are first unmasked and then divided by $\boldsymbol{q}(x)$. An array of syndrome-trellis decoders then searches the corresponding candidate cosets of $H$ for the most likely error vector. Among all decoded candidates, the one with minimum Hamming weight is selected, and the plaintext is finally recovered by applying the inverse permutation $P^{-1}$.The complete decryption procedure is presented in Section 5.

## 3. ENCRYPTION BY SENDER

Assuming the sender possesses the public key, encryption proceeds as follows.

**Step1-Plaintext Generation:**
A random error vector

$$\boldsymbol{e} \in \mathbb{F}_2^N$$

is generatedby independently flipping each bit of an all-zero sequence of length $N$ with probability $\varepsilon$.

**Step 2–Syndrome Computation:**

The ciphertext, represented by the syndrome

$$\boldsymbol{s} \in \mathbb{F}_2^{N-K}$$

is computed as

$$\boldsymbol{s} = \widehat{H}\boldsymbol{e}^T. \quad (22)$$

**Step 3-Transmission:**

The syndrome $\boldsymbol{s}$is transmitted to the recipient.

In practice, rare decoding failures can be detected reliably either by appending a CRC during encryption or by verifying that the Hamming weight of the decoded error vector matches its expected value (see Section 5). Upon detection of a failure, retransmission may be requested.

## 4. CONSTRUCTION OF SYNDROME TRELLISES

We now describe the construction of a directed graph representing a coset of a CC. The procedure is most easily illustrated by the example introduced in Section 2 with

$$H(x) = [1+x^2,\ 1+x+x^2]$$

The corresponding scalar parity-check matrix $H$, given by Equation (11), is

$$H = \begin{bmatrix} 11 & & & & \\ \mathbf{01} & 11 & & & \\ \mathbf{11} & 01 & 11 & \cdots & \\ & 11 & 01 & & 11 \\ & & 11 & & 01 \\ & & & & 11 \end{bmatrix}$$

Let

$$\boldsymbol{s} = (s_0, s_1, s_2, \dots, s_{N-K-1})^{\mathrm{T}}$$

denote the syndrome vector, and let

$$\boldsymbol{e} = (e_0, e_1, e_2, \dots, e_{N-1})$$

denote the error vector. Each syndrome bit is obtained as the modulo-2 sum of the columns of $H$ corresponding to positions where$e_i = 1$. Suppose first that $s_0$ = 0. The two possible pairs of error bits that generate this syndrome bit are

$$(e_0, e_1) = 00 \text{ or} (e_0, e_1) = 11$$

These error pairs determine not only the syndrome bit $s_0$, but also contribute to the subsequent syndrome bits $s_1$and $s_2$through the corresponding submatrix

$$\tau = \begin{bmatrix} 01 \\ 11 \end{bmatrix}$$

shown in bold in $H$ above. We refer to the linear combinations of the columns of $\tau$ as the **states**of the syndrome trellis. For the pair $(e_0, e_1) = 00$, the resulting state is $[00]^T$, whereas for$(e_0, e_1) = 11$,it is$[10]^T$. Similarly, when$s_0$ = 1 the admissible error pairs become

$$(e_0, e_1) = 01 \text{ or} (e_0, e_1) = 10$$

yielding the states $[11]^T$ and $[01]^T$, respectively. Thus, the syndrome trellis has two possible initial configurations, depending on the value of the syndrome bit. For $s_0$ = 0, the trellis begins as

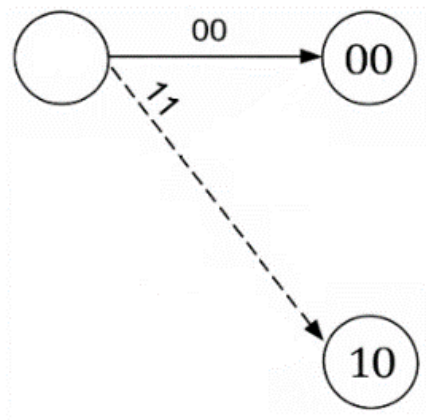

while for $s_0$ = 1 it begins as

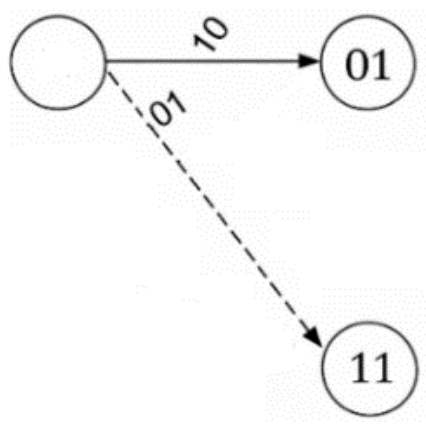

Consider now the $i$-thsyndrome bit $s_i$.Its value depends not only on the corresponding pair of error bits $(e_{2i}, e_{2i+1})$, but also on the source statefrom which the trellis transition originates. The syndrome-trellis segments corresponding to $s_i$ = 0 for each of the four possible source states, are shown in Fig.1.

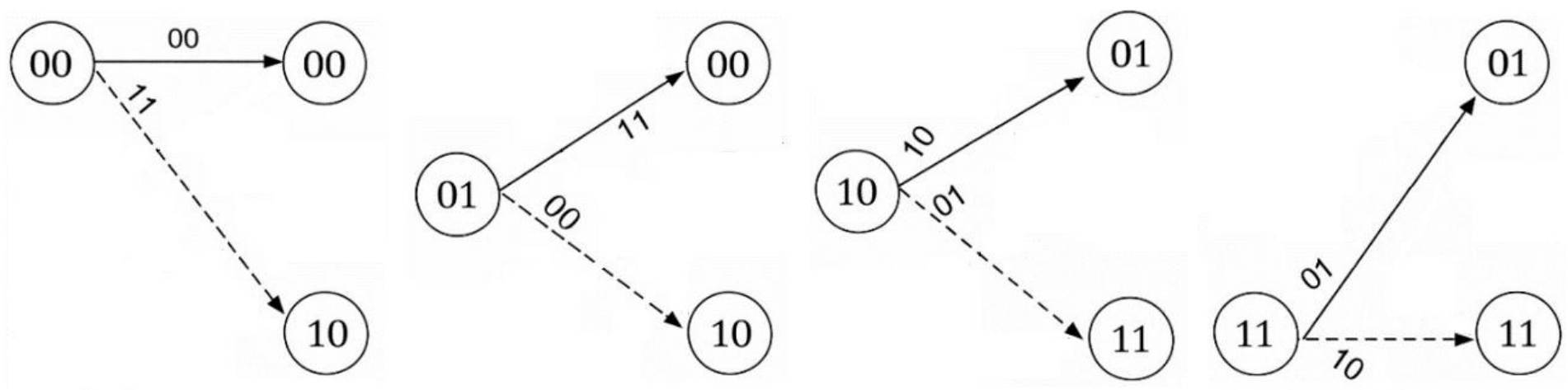

(a) (b) (c) (d)

Figure 1. Syndrome trellis segments corresponding to $s_i$ = 0 andsource states: (a) $[00]^T$, (b) $[01]^T$, (c) $[10]^T$ and (d) $[11]^T$.

Similarly, the syndrome-trellis segments corresponding to $s_i$ = 1, for each of the four possible source states, are shownin Fig.2.

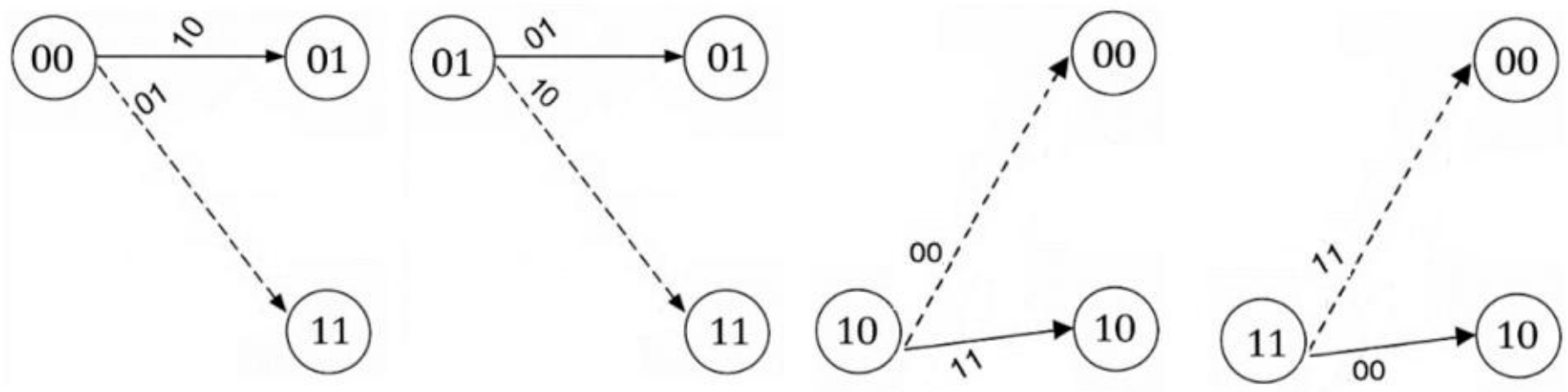

(a) (b) (c) (d)
Figure 2. Syndrome trellis segments corresponding to $s_i$ = 1 and source states: (a) $[00]^T$, (b) $[01]^T$, (c) $[10]^T$ and (d) $[11]^T$.

These segments define two building blocks ofthe syndrome trellis: one corresponds to $s_i$ = 0 and the other to $s_i$ = 1, as illustrated in Fig. 3

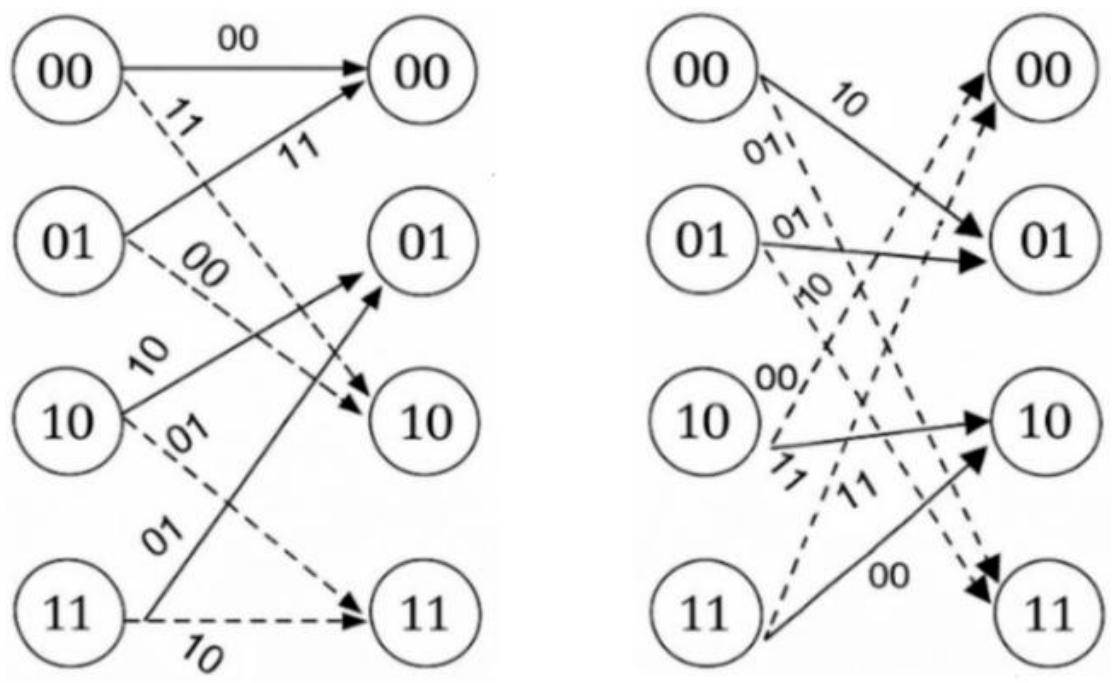

(a) (b)
Figure 3. Syndrome-trellisbuilding blocks corresponding to: (a) $s_i = 0$, (b) $s_i = 1$.

Since both building blocks share the same state space, they can be interconnected through their common states, allowing the syndrome trellis for any coset of the CC to be generated dynamically according to the syndrome values. The trellis always starts from the state $[00\ldots0]^T$ and terminates at a unique state determined by the $p$ tail syndrome bits. In the present example, $p$=2.

**Branch Metrics of Syndrome Trellises:** The metric assigned to each trellis branch is simply its Hamming weight, namely the number of ones in the corresponding pair of error bits.

**Maximum-Likelihood Decoding of Syndrome Trellises:**

Decoding consists of searching the syndrome trellis for the minimum-Hamming-weight path. The procedure is as follows.

1. **Initialization:**At time 0, the state$[00...0]^T$is assigned metric 0, while all other states are assigned a sufficiently large positive value to prevent their selected during the search.
2. **Branch Metric Assignment:**Assigneach trellis branch a metric equal to the Hamming weight of the corresponding error vector segment.
3. **State Metric Computation:**For each state, compute the accumulated metric of the minimum-weight path terminating at that state.

4. **Survivor Selection:** For each state, retain only the incoming branch belonging to the minimum-weight path determined in Step 3 and discard the others. The retained branch is referred to as the *survivor branch*.
5. **Trace back:** Starting from the terminal state, trace back through the survivor branches to the initial state, thereby reconstructing the minimum-weight error vector $\hat{\boldsymbol{e}}$.

## 5. DECRYPTION BY THE RECIPIENT

The recipient possesses the public and private keys together with the received cipher text:

$$\{H, \widetilde{H}, \widehat{H}, S, P, \boldsymbol{q}(x), \varepsilon, \boldsymbol{s}\}$$

Decryption proceeds as follows.

**Step 1–Inversion of the Non singular transformation**

The inverse transformation $S^{-1}$is applied to the received syndrome

$$\tilde{\boldsymbol{s}} = S^{-1}\boldsymbol{s} \tag{23}$$

**Step 2–Unmasking**

The transformed syndrome can be written as

$$\tilde{\boldsymbol{s}} = \left(H' \oplus \widetilde{H}\right)P\boldsymbol{e}^T = H'P\boldsymbol{e}^T \oplus \widetilde{H}P\boldsymbol{e}^T = H'P\boldsymbol{e}^T \oplus \boldsymbol{l}_i, \boldsymbol{l}_i \in \mathrm{LS}(\mathcal{L}) \tag{24}$$

Since $\widetilde{H}P$has the same column span as $\widetilde{H}$, the masking contribution $\boldsymbol{l}_i$belongs to $\mathrm{LS}(\mathcal{L})$. Although$P$ and$\widetilde{H}$areknown to the recipient, the particular masking vector $\boldsymbol{l}_i$is unknown and may assume any of the $2^l$values in $\mathrm{LS}(\mathcal{L})$. Consequently, the set of candidate unmasked syndromes is

$$\mathfrak{D}' = \{\tilde{\boldsymbol{s}} \oplus \boldsymbol{l}_i \mid \boldsymbol{l}_i \in \mathrm{LS}(\mathcal{L})\} \tag{25}$$

**Step 3–Division by the High-Memory Polynomial**

The memory of the CC is then reduced by dividing each candidate unmasked syndrome in $\mathfrak{D}'$bythe high-memory polynomial $\boldsymbol{q}(x)$, yielding the set of "low-memory" candidate syndromes

$$\mathfrak{D} = \left\{\frac{\tilde{s} \oplus \boldsymbol{l}_i}{\boldsymbol{q}(\boldsymbol{x})} \mid \boldsymbol{l}_i \in \mathrm{LS}(\mathcal{L})\right\} \tag{26}$$

If no transmission errors have occurred and the correct masking was removed at Step 2,the division by $\boldsymbol{q}(x)$produceszero remainder.

**Step 4–Parallel Syndrome-Trellis Decoding**

The objective is to determine the most likely value of$P\boldsymbol{e}^T$ by decoding all members of$\mathfrak{D}$. Since each member of $\mathfrak{D}$ defines a distinct coset of the CC, it also defines a distinct syndrome trellis. because

$$|\mathfrak{D}| = 2^l$$

all candidate trellises may be decoded in parallel. The estimated error vector $P\hat{\boldsymbol{e}}^T$is then selected as the minimum-weight solution among the$2^l$decoded candidates.

A key distinction exists between the candidate low-memory syndromes. If the correct masking vector was removed in Step 2, decoding yields an error vector satisfying

$$\mathrm{wt}(P\hat{\boldsymbol{e}}^T) \approx \varepsilon N \tag{27}$$

In contrast, removing an incorrect masking vector effectively injects random errors into the syndrome, shifting the corresponding coset to an essentially arbitrary one. Consequently, the decoded error vectors $P\tilde{\boldsymbol{e}}^T$ associated with incorrect candidates typically exhibit substantially larger weight,

$$\mathrm{wt}(P\tilde{\boldsymbol{e}}^T) \approx \delta \tag{28}$$

where $\delta$ is the typical nearest-neighbor distance defined by Equation (19). This separation in Hamming weight enables reliable identification of the correct plaintext.

**Step 5 - Plaintext Recovery**

Since $P$ is a permutation matrix,

$$P^{-1} = P^T$$

The plaintext is therefore recovered as

$$\hat{\boldsymbol{e}}^T = P^T P\hat{\boldsymbol{e}}^T \tag{29}$$

If a CRC was appended to the plaintext during encryption, it is verified at this stage. If the CRC check fails, the next most likely decoded candidate is examined. The process continues until a valid plaintext is identified or all candidates have been exhausted, in which case retransmission may be requested.

The complete encryption and decryption procedures are summarized in the block diagram shown in Fig. 4.

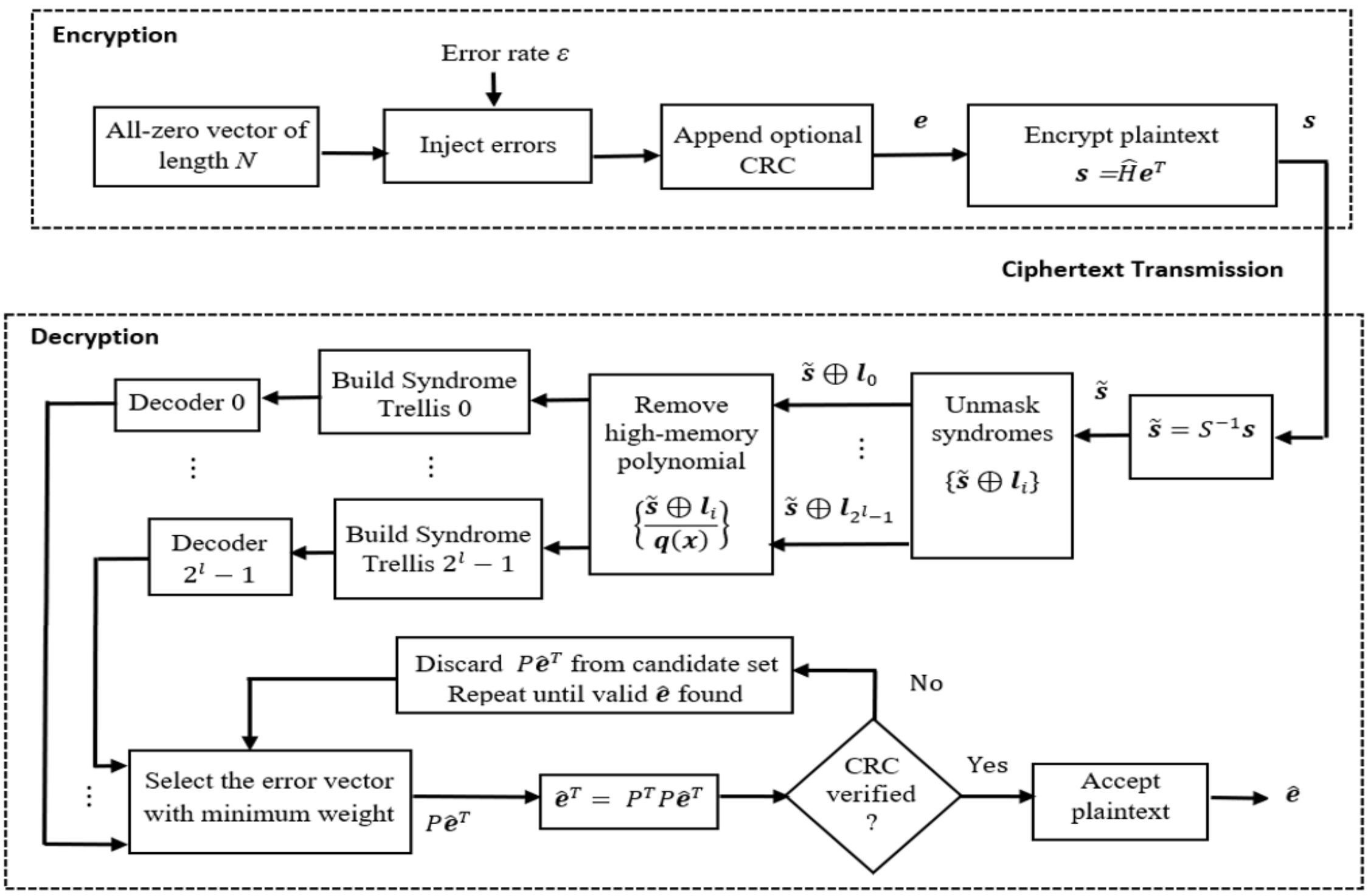

Figure 4. Encryption and Decryption block diagram.

## 6. WORKED EXAMPLE

Consider the CC defined by

$$H(x) = [1+x^2, 1+x+x^2]$$

and select the high-memory polynomial

$$\boldsymbol{q}(x) = 1+x^3$$

The resulting high-memory parity-check matrix is

$$H'(x) = [1+x^2+x^3+x^5,\ 1+x+x^2+x^3+x^4+x^5]$$

Assume aplaintext of length 16, with no CRC appended. The corresponding finite-dimensional matrix $H'$ is of size13×16

$$H' = \begin{bmatrix} 1&1&0&0&0&0&0&0&0&0&0&0&0&0&0&0 \\ 0&1&1&1&0&0&0&0&0&0&0&0&0&0&0&0 \\ 1&1&0&1&1&1&0&0&0&0&0&0&0&0&0&0 \\ 1&1&1&1&0&1&1&1&0&0&0&0&0&0&0&0 \\ 0&1&1&1&1&1&0&1&1&1&0&0&0&0&0&0 \\ 1&1&0&1&1&1&1&1&0&1&1&1&0&0&0&0 \\ 0&0&1&1&0&1&1&1&1&1&0&1&1&1&0&0 \\ 0&0&0&0&1&1&0&1&1&1&1&1&0&1&1&1 \\ 0&0&0&0&0&0&1&1&0&1&1&1&1&1&0&1 \\ 0&0&0&0&0&0&0&0&1&1&0&1&1&1&1&1 \\ 0&0&0&0&0&0&0&0&0&0&1&1&0&1&1&1 \\ 0&0&0&0&0&0&0&0&0&0&0&0&1&1&0&1 \\ 0&0&0&0&0&0&0&0&0&0&0&0&0&0&1&1 \end{bmatrix}$$

We now construct the masking matrix$\widetilde{H}$by selecting its columns randomly from the linear spanLS($\mathcal{L}$), where $\mathcal{L}$ is a set of $l$ random vectors of length 13. Although $\mathcal{L}$may contain arbitrary random vectors of length 13, for simplicity we choose two alternating patterns with Hamming weight approximately equal to$(N-K)/2$, thereby ensuring statistical balance:

$$\mathcal{L} = \{(1010101010101)^{\mathrm{T}}, (0101010101010)^{\mathrm{T}}\}$$

The corresponding linear span is therefore

$$\mathrm{LS}(\mathcal{L}) = \{\mathbf{0}^{\mathrm{T}}, (0101010101010)^{\mathrm{T}}, (1010101010101)^{\mathrm{T}}, \mathbf{1}^{\mathrm{T}}\}$$

where **0** and **1**denote the all-zero and all-one vectors, respectively. A dense masking matrix $\widetilde{H}$is then formed by selecting each of its columns independently fromLS($\mathcal{L}$)

$$\widetilde{H} = \begin{bmatrix}
1&0&0&1&1&1&0&0&1&0&0&1&1&1&0&1\\
0&1&0&1&0&0&1&1&0&1&1&0&0&0&1&0\\
1&0&0&1&1&1&0&0&1&0&0&1&1&1&0&1\\
0&1&0&1&0&0&1&1&0&1&1&0&0&0&1&0\\
1&0&0&1&1&1&0&0&1&0&0&1&1&1&0&1\\
0&1&0&1&0&0&1&1&0&1&1&0&0&0&1&0\\
1&0&0&1&1&1&0&0&1&0&0&1&1&1&0&1\\
0&1&0&1&0&0&1&1&0&1&1&0&0&0&1&0\\
1&0&0&1&1&1&0&0&1&0&0&1&1&1&0&1\\
0&1&0&1&0&0&1&1&0&1&1&0&0&0&1&0\\
1&0&0&1&1&1&0&0&1&0&0&1&1&1&0&1\\
0&1&0&1&0&0&1&1&0&1&1&0&0&0&1&0\\
1&0&0&1&1&1&0&0&1&0&0&1&1&1&0&1
\end{bmatrix}.$$

The masked parity-check matrix is obtained as

$$H' \oplus \widetilde{H} = \begin{bmatrix}
0&1&0&1&1&1&0&0&1&0&0&1&1&1&0&1\\
0&0&1&0&0&0&1&1&0&1&1&0&0&0&1&0\\
0&1&0&0&0&0&0&0&1&0&0&1&1&1&0&1\\
1&0&1&0&0&1&0&0&0&1&1&0&0&0&1&0\\
1&1&1&0&0&0&0&1&0&1&0&1&1&1&0&1\\
1&0&0&0&1&1&0&0&0&0&0&1&0&0&1&0\\
1&0&1&0&1&0&1&1&0&1&0&0&0&0&0&1\\
0&1&0&1&1&1&1&0&1&0&0&1&0&1&0&1\\
1&0&0&1&1&1&1&1&1&1&1&0&0&0&0&0\\
0&1&0&1&0&0&1&1&1&0&1&1&1&1&0&1\\
1&0&0&1&1&1&0&0&1&0&1&0&1&0&1&0\\
0&1&0&1&0&0&1&1&0&1&1&0&1&1&1&1\\
1&0&0&1&1&1&0&0&1&0&0&1&1&1&1&0
\end{bmatrix}.$$

The resulting matrix is dense, with approximately half of its entries are equal to 1, while the row and column weights are concentrated around $(N-K)/2$ and $N/2$, respectively. Consequently, the original diagonal structure of $H$is effectively concealed.

Next, a random column permutation is applied using a 16×16 permutation matrix $P$

$$P = \begin{bmatrix}
0&0&0&0&0&0&0&0&0&0&0&0&0&1&0&0\\
0&0&0&0&0&0&0&0&0&0&0&0&0&0&1&0\\
0&0&0&0&0&0&0&0&1&0&0&0&0&0&0&0\\
0&0&0&0&0&0&0&1&0&0&0&0&0&0&0&0\\
0&0&0&0&0&0&0&0&0&1&0&0&0&0&0&0\\
0&0&0&0&0&0&0&0&0&0&1&0&0&0&0&0\\
1&0&0&0&0&0&0&0&0&0&0&0&0&0&0&0\\
0&0&0&1&0&0&0&0&0&0&0&0&0&0&0&0\\
0&0&0&0&0&0&0&0&0&0&0&1&0&0&0&0\\
0&0&1&0&0&0&0&0&0&0&0&0&0&0&0&0\\
0&0&0&0&1&0&0&0&0&0&0&0&0&0&0&0\\
0&0&0&0&0&1&0&0&0&0&0&0&0&0&0&0\\
0&1&0&0&0&0&0&0&0&0&0&0&0&0&0&0\\
0&0&0&0&0&0&0&0&0&0&0&0&1&0&0&0\\
0&0&0&0&0&0&1&0&0&0&0&0&0&0&0&0\\
0&0&0&0&0&0&0&0&0&0&0&0&0&0&0&1
\end{bmatrix}$$

Applying this permutation to the masked matrix produces

$$(H' \oplus \widetilde{H})P = \begin{bmatrix}
0&1&0&0&0&1&0&1&0&1&1&1&1&0&1&1\\
1&0&1&1&1&0&1&0&1&0&0&0&0&0&0&0\\
0&1&0&0&0&1&0&0&0&0&0&1&1&0&1&1\\
0&0&1&0&1&0&1&0&1&0&1&0&0&1&0&0\\
0&1&1&1&0&1&0&0&1&0&0&0&1&1&1&1\\
0&0&0&0&0&1&1&0&0&1&1&0&0&1&0&0\\
1&0&1&1&0&0&0&0&1&1&0&0&0&1&0&1\\
1&0&0&0&0&1&0&1&0&1&1&1&1&0&1&1\\
1&0&1&1&1&0&0&1&0&1&1&1&0&1&0&0\\
1&1&0&1&1&1&0&1&0&0&0&1&1&0&1&1\\
0&1&0&0&1&0&1&1&0&1&1&1&0&1&0&0\\
1&1&1&1&1&0&1&1&0&0&0&0&1&0&1&1\\
0&1&0&0&0&1&1&1&0&1&1&1&1&1&0&0
\end{bmatrix}$$

The permutation destroys the remaining positional ordering of the columns, further concealing the original convolutional structure.We then select a nonsingular matrix $S$, together with its inverse $S^{-1}$

$$S=\begin{bmatrix}
1&1&0&0&0&0&0&0&0&0&0&0&0\\
0&1&1&0&0&0&0&0&0&0&0&0&0\\
0&0&1&1&0&0&0&0&0&0&0&0&0\\
0&0&0&1&1&0&0&0&0&0&0&0&0\\
0&0&0&0&1&1&0&0&0&0&0&0&0\\
0&0&0&0&0&1&1&0&0&0&0&0&0\\
0&0&0&0&0&0&1&1&0&0&0&0&0\\
0&0&0&0&0&0&0&1&1&0&0&0&0\\
0&0&0&0&0&0&0&0&1&1&0&0&0\\
0&0&0&0&0&0&0&0&0&1&1&0&0\\
0&0&0&0&0&0&0&0&0&0&1&1&0\\
0&0&0&0&0&0&0&0&0&0&0&1&1\\
0&0&0&0&0&0&0&0&0&0&0&0&1
\end{bmatrix}, S^{-1}=\begin{bmatrix}
1&1&1&1&1&1&1&1&1&1&1&1&1\\
0&1&1&1&1&1&1&1&1&1&1&1&1\\
0&0&1&1&1&1&1&1&1&1&1&1&1\\
0&0&0&1&1&1&1&1&1&1&1&1&1\\
0&0&0&0&1&1&1&1&1&1&1&1&1\\
0&0&0&0&0&1&1&1&1&1&1&1&1\\
0&0&0&0&0&0&1&1&1&1&1&1&1\\
0&0&0&0&0&0&0&1&1&1&1&1&1\\
0&0&0&0&0&0&0&0&1&1&1&1&1\\
0&0&0&0&0&0&0&0&0&1&1&1&1\\
0&0&0&0&0&0&0&0&0&0&1&1&1\\
0&0&0&0&0&0&0&0&0&0&0&1&1\\
0&0&0&0&0&0&0&0&0&0&0&0&1
\end{bmatrix}$$

and construct the final public matrix

$$\widehat{H} = S(H' \oplus \widetilde{H})P = \begin{bmatrix}
1&1&1&1&1&1&1&1&1&1&1&1&1&0&1&1\\
1&1&1&1&1&1&1&0&1&0&0&1&1&0&1&1\\
0&1&1&0&1&0&0&0&1&0&1&1&1&1&1&1\\
0&1&0&1&1&1&1&0&0&0&1&0&1&0&1&1\\
0&1&1&1&0&0&1&0&1&1&1&0&1&0&1&1\\
1&0&1&1&0&1&1&0&1&0&1&0&0&0&0&1\\
0&0&1&1&0&1&0&1&1&0&1&1&1&1&1&0\\
0&0&1&1&1&1&0&0&0&0&0&0&1&1&1&1\\
0&1&1&0&0&1&0&0&0&1&1&0&1&1&1&1\\
1&0&0&1&0&1&1&0&0&1&1&0&1&1&1&1\\
1&0&1&1&0&0&0&0&0&1&1&1&1&1&1&1\\
1&0&1&1&1&1&0&0&0&1&1&1&0&1&1&1\\
0&1&0&0&0&1&1&1&0&1&1&1&1&1&0&0
\end{bmatrix}$$

The resulting matrix $\widehat{H}$is dense and exhibits strong random-like statistical properties, with no visible diagonal or convolutional structure remaining. Since $S$ is non singular and $P$ is a permutation matrix, both transformations preserve matrix rank. Consequently, if $H \oplus \widetilde{H}$ has full rank, then the public matrix $\widehat{H}$also has full rank.

**Encryption**:
Consider the plain text error vector

$$\boldsymbol{e} = [0\ 0\ 0\ 0\ 0\ 0\ 0\ 1\ 0\ 0\ 0\ 0\ 0\ 0\ 0\ 0]$$

which is a unique coset leader due to its low weight and error distribution. The cipher text syndrome is computed as

$$\boldsymbol{s} = \widehat{H}\boldsymbol{e}^T = [1000001000001]^T.$$

**Decryption:**

**Step1- Inverse Row Operations**
Apply the inverse row transformation to the received syndrome:

$$\tilde{\boldsymbol{s}} = S^{-1}\boldsymbol{s} = [1000000111111]^T$$

**Step 2 - Unmasking**

For the selected $\mathrm{LS}(\mathcal{L})$, the set $\mathfrak{D}'$of candidate unmasked syndromes is

$$\mathfrak{D}' = \{[1\ 0\ 0\ 0\ 0\ 0\ 0\ 1\ 1\ 1\ 1\ 1\ 1]^T, [1\ 1\ 0\ 1\ 0\ 100101\ 0\ 1]^T, [0\ 0\ 1\ 0\ 1\ 011010\ 1\ 0]^T, [0\ 1\ 1\ 1\ 1\ 110000\ 0\ 0]^T\}$$

**Step 3–Division by $\boldsymbol{q}(x)$**

The set $\mathfrak{D}$of low-memory candidate syndromes is obtained by dividing each member of $\mathfrak{D}'$by $\boldsymbol{q}(x) = 1+x^3$. This yields four low-memory candidate syndromes, of which only one is guaranteed to be exactly divisible by$\boldsymbol{q}(x)$.

$\boldsymbol{s}_0$=[0 00 0 0 00111]$^T$, remainder1,
$\boldsymbol{s}_1$=[0101111101]$^T$, remainder 1,
$\boldsymbol{s}_2$=[0010111010]$^T$,
$\boldsymbol{s}_3$= [0111000000]$^T$.

Thus, not all divisions yield zero remainder. Assuming no transmission errors have occurred, only candidates with zero remainder are consistent with exact division by $\boldsymbol{q}(x)$and may therefore be retained for syndrome-trellis decoding. In the general case, however, all candidates must be considered, since transmission errors may perturb the syndrome and produce nonzero remainders even for the correct masking vector.

**Step 4- Parallel Syndrome-Trellis Decoding**

The candidate syndrome trellises are decoded in parallel to determine the most likely error vector$P\hat{\boldsymbol{e}}^T$. In this example, the CC has a free distance $d_{free} = 5$, allowing reliable correction of up to two errors within a sliding window of six bits. If three or more errors occur within a window, decoding may fail.

The syndrome trellis corresponding to$\boldsymbol{s}_3$=[0 1 1 10 0 0 0 0 0]$^T$is illustrated in Fig. 5. Since $p = 2$, the final two syndrome bits uniquely determine the terminal state of the syndrome trellis.

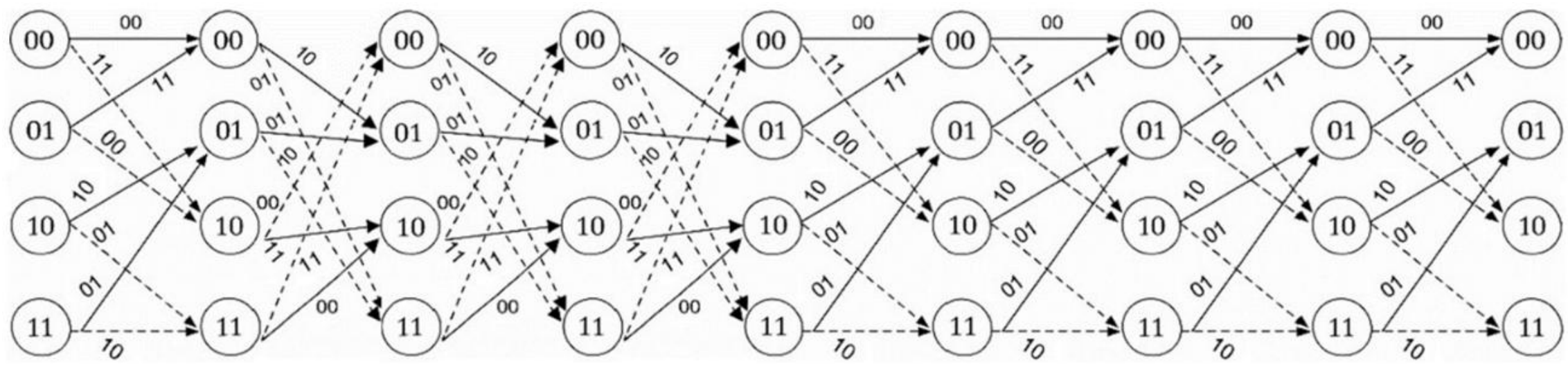

Figure 5. Syndrome trellis corresponding to $\boldsymbol{s}_3$= [0 1 1 10 000 0 0]$^T$

Fig. 6shows the trellis of Fig. 5after enforcing unique initial and terminal states. The highlighted path corresponds to the minimum-weight path, yielding

$$P\hat{\boldsymbol{e}}^T = [0\ 0\ 0\ 1\ 0\ 0\ 0\ 0\ 0\ 0\ 0\ 0\ 0\ 0\ 0\ 0]^T$$

The selected path has Hamming weight 1. By contrast, the minimum-weight paths in all remaining candidate syndrome trellises have weights greater than 1.

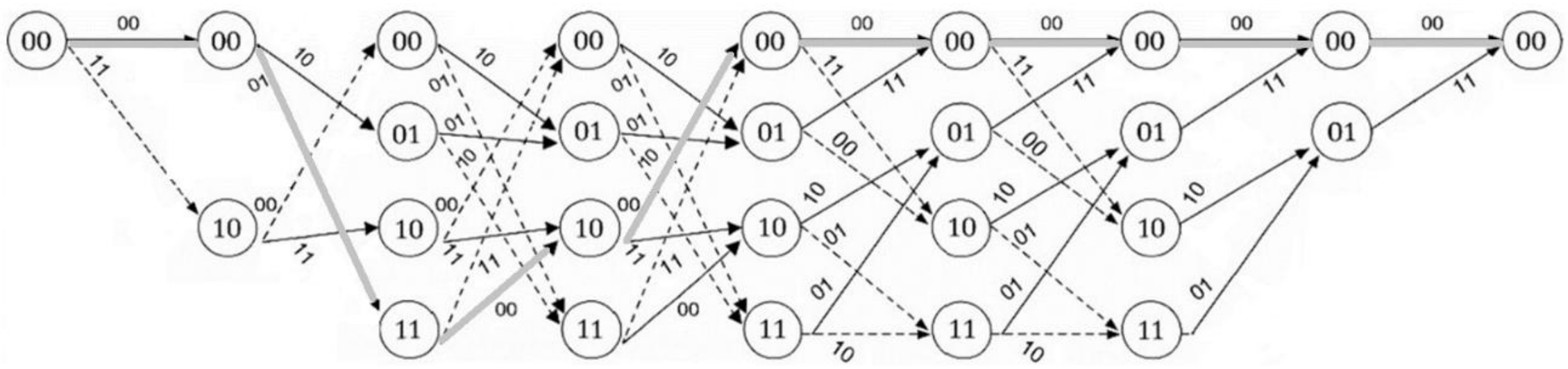

Figure 6. Minimum-weight path through the syndrome trellis of Fig. 5

**Step 5- Plaintext Recovery**

The plaintext is recovered by applying the inverse permutation:

$$\hat{\boldsymbol{e}}^T = P^T P\hat{\boldsymbol{e}}^T = [0\ 0\ 0\ 0\ 0\ 0\ 0\ 1\ 0\ 0\ 0\ 0\ 0\ 0\ 0\ 0]^T$$

The recovered vector matches original plaintext.

## 7. Cryptanalysis Resistance and Performance

**Threat Model**: We consider a passive adversary (eavesdropper) who has full knowledge of the public key $\widehat{H}$, the encryption algorithm, and all system parameters except the private key. The adversary observes the intercepted ciphertext (syndrome) $\boldsymbol{s}$ and seeks either to recover the plaintext error vector $\boldsymbol{e}$ or to recover the hidden structure of the public key.Direct recovery of $\boldsymbol{e}$ by inverting $\widehat{H}$is computationally infeasible because the corresponding syndrome defines a coset containing an astronomically large number of candidate error vectors.

We further assume that, for the selected parameters, the public matrix $\widehat{H}$is computationally indistinguishable from the parity-check matrix of a random binary linear code. As in the Classic McEliece cryptosystem, this assumption is not supported by a formal proof but by the absence of any known efficient structural attacks against appropriately chosen parameter sets. Under these assumptions, the adversary is effectively limited to three classes of attacks:

**Plaintext-recovery attacks**, which attempt to recover the transmitted error vector by solving the syndrome decoding problem. The complexity of these attacks grows exponentially with the error weight, which is itself unknown in the proposed framework because only the error probability $\varepsilon$ is public.

**Structure-recovery attacks**, which attempt to recover the hidden structured matrix $H$, together with the masking and transformation operations, from the public matrix

**Distinguishing attacks**, whose objective is to determine whether $\widehat{H}$ can be distinguished from the parity-check matrix of a random linear code.

As in the Classic McEliece cryptosystem, this assumption is not supported by a formal proof but rather by the absence of any known efficient structural attack against appropriately chosen parameter sets.

The second and third attack classes are hindered by three complementary obfuscation mechanisms:

- **Dense masking.** The memory of the CCis substantially increased by multiplication with a high-degree polynomial, followed by masking with $\widetilde{H}$, whose columns are statistically balanced random vectors with Hamming weight close to $(N-K)/2$. This process effectively conceals the underlying algebraic structure.
- **Random row operations.** A nonsingular matrix $S$randomizes the basis of the masked parity-check matrix through invertible linear combinations of its rows.
- **Random column permutation.** A permutation matrix$P$ uniformly permutes the coordinate positions, eliminating any remaining positional or diagonal patterns.

Additional analytical evidence that the resulting public matrix exhibits the statistical properties expected of a random full-rank $(N,K)$ binary code is presented in [8].

Although the syndrome $\boldsymbol{s}$ contains no explicitly injected errors, the decryption problem associated with the proposed masked parity-check matrices still appears to reduce to the classical syndrome decoding problem, which is NP-hard. Daniel J. Bernstein [17] analyzed quantum speedups for information-set decoding (ISD) attacks. His analysis addresses the decoding problem itself, independently of whether the public code is represented by a generator matrix or a parity-check matrix. Consequently, provided that ISD remains the best-known generic attack against the proposed framework, Bernstein's complexity estimates apply directly using the effective code parameters defined by the public parity-check matrix.

In the proposed framework, given a finite-dimensional masked parity-check matrix $\widehat{H}$anda syndrome $\boldsymbol{s}$, and assuming that exactly $\varepsilon N$ errors occurred, the attacker must recover an error vector $\boldsymbol{e}$ satisfying

$$\widehat{H}\boldsymbol{e}^T = \boldsymbol{s},\ \mathrm{wt}(\boldsymbol{e}) = \varepsilon N \tag{30}$$

The cryptanalytic complexity is estimated using the best-known information-set decoding (ISD) algorithms, including those of Prange, Lee–Brickell, and BJMM [14–16], which measure security primarily in terms of bit operations and typically do not account for memory requirements or large-scale parallelization. As with other code-based cryptosystems, the proposed construction relies on sufficiently large parameter sets; smaller parameters may permit exhaustive search or structural attacks and are therefore not intended to provide cryptographic security. Accordingly, we consider only parameter ranges for which generic ISD attacks are believed to dominate.

The analyses in [14–16] generally assume that the Hamming weight wt($\boldsymbol{e}$) is known a priori. In the proposed framework, however, the exact number of errors is not fixed but is determined probabilistically by the publicly known error probability $\varepsilon$. This additional uncertainty further complicates generic decoding attacks. Moreover, since the error vector $\boldsymbol{e}$ is generated as a statistically random binary vector, it contains no intentional algebraic or combinatorial structure that could be exploited to reduce decoding complexity.

In the context of quantum computing, ISD complexity estimates are typically derived from classical attack models. Although quantum algorithms such as Grover's search algorithm [17] provide a quadratic speedup, no exponential quantum speedup for ISD is currently known. Consequently, code-based cryptography generally assumes that ISD, with at most quadratic quantum acceleration, remains the most efficient generic attack.

To compare the proposed framework with other code-based cryptosystems, we adopt the complexity measure of [17]. Specifically, for a classical attack on an $(N,K,t)$ code, the ISD complexity, denoted by $C_{\mathrm{ISD}}$, is proportional to the number of iterations required by the ISD algorithm. In the present work, the classical ISD complexity is approximated by

$$C_{\mathrm{ISD}} \sim \frac{\binom{N}{K}}{0.29\binom{N-t}{K}} \tag{31}$$

For quantum computing, the corresponding complexity, denoted by $C_{\mathrm{QISD}}$, becomes

$$C_{\mathrm{QISD}} \sim \sqrt{\frac{\binom{N}{K}}{0.29\binom{N-t}{K}}} \tag{32}$$

Each quantum iteration requires a function evaluation requiring $O(N^3)$ qubit operations.

We now apply Equation (31) to estimate the ISD complexity of a Goppa-code-based cryptosystem, using the widely adopted parameter set $(N, K,t)$ = (8192,6528,128) as a reference, and compare it with the proposed MPC framework at a comparable key size(~50 Mbits). Public-key size is measured as the total number of stored bits required to represent the dense binary parity-check matrix without compression. Although the parameter $\varepsilon$ is publicly known, the actual number of errors in the plaintext vector $\boldsymbol{e}$ is not revealed. For a conservative comparison, however, we assume that the effective number of errors in the MPC scheme is fixed at $\varepsilon N$. Substituting the parameters of the (8192, 6528, 128) Goppa code into Equation (31) yields

$$C_{\mathrm{ISD}}(\text{Goppa}) \approx \frac{\binom{8192}{6528}}{0.29\binom{8192-128}{6528}} \approx 2.74\times10^{84} \tag{33}$$

For an MPCconstruction with a comparable key size (~35 Mbits), we may choose $N$ =8200 and $N-K$= 4300. Selecting the degree-14 CC polynomials

$$\boldsymbol{p}_0(x)= 55751,\ \boldsymbol{p}_1(x)= 66323, \tag{34}$$

expressed in octal form, yields a code with $d_{free}$= 20. Choosing $\boldsymbol{q}(x)$of degree 186gives$p + q$ = 200. For these parameters, selecting $\varepsilon$ = 0.07producesapproximately 574 errors in an8200-bit error vector.This error rate lieswell within the correction capability of the selected CC, which corrects up to 9 errors in a 30-bit window, making decoding failures highly unlikely.

By contrast, incorrectly unmasked candidates are decoded, asdiscussed in Section 2, to vectors lying at a Hamming distance of approximately

$$0.11\times 8200 \approx 902$$

from the received vector. This distance is substantially larger than the expected 574-error radius of the valid plaintext, placing such candidates well outside the decoder's effective decoding radius. This separation enables reliable discrimination between the correct plaintext and incorrect candidates. Consequently, the valid plaintext is identified with overwhelming probability, while incorrect candidates are systematically rejected.

Under these assumptions, the estimated ISD complexity of the proposed MPC cryptosystem is

$$C_{\text{ISD}}(\text{MPC}) \approx \frac{\binom{N}{K}}{0.29\binom{N-\varepsilon N}{K}} = \frac{\binom{8200}{4300}}{0.29\binom{8200-574}{4300}} \approx 1.44\times 10^{196} \quad (35)$$

This corresponds to an estimated increase in ISD attack complexity by a factor of approximately

$$\frac{C_{\text{ISD}}(\text{MPC})}{C_{\text{ISD}}(\text{Goppa})} \approx 0.52\times 10^{112} \approx 2^{371}$$

relative to the Niederreiter and Classic McEliece parameter set considered in Equation (33). For quantum ISD (QISD),the improvement is approximately $2^{185}$, consistent with the square-root speedup provided by Grover-typequantum search.

The probability of decoding failure, and hence the need for retransmission, can also be estimated directly. For example, the decoder fails if more than 9 errors occur within a 30-bit window. At an error probability of $\varepsilon$ = 0.07, the probability of this event is approximately

$$1.6\times 10^{-6}$$

Over 274 windows, corresponding to a ciphertext length of approximately 8,200 bits, the probability of successful decoding is therefore

$$(1 - 1.6\times 10^{-6})^{274} \approx 0.9995$$

This demonstrates that the proposed MPC cryptosystem maintains a very high probability of successful decryption even for ciphertexts spanning thousands of bits under fully random error injection.

We note that low-rate public codes can correct a larger fraction of errors than high-rate codes. However, their increased redundancy may adversely affect both efficiency and security in code-based cryptosystems, as indicated by (31). When $K \ll N$, the large number of parity constraints, $N-K$, may be exploited to accelerate decoding attacks [18]. Consequently, the code rate is an important design parameter in selecting the underlying CC. In practice, a code rate close to 1/2, i.e., $N-K \approx K$, provides a good balance between security and implementation efficiency.

Finally, we examine the computational complexity of decryption. Schöffel et al. [19] presented a hardware/software co-design implementation of the Hamming Quasi-Cyclic (HQC) cryptosystem for IoT edge devices, demonstrating that code-based cryptography can provide a practical alternative to lattice-based schemes in resource-constrained environments. Similarly, [20] introduced an FPGA-based hardware accelerator for HQC post-quantum cryptography, implementing key generation, encapsulation, and decapsulation with substantial performance gains through tight integration with a RISC-V core.

In the proposed MPC cryptosystem, the dominant contribution to the decryption complexity arises from syndrome-trellis decoding. Other operations, such as polynomial division and syndrome unmasking, are performed on the entire ciphertext but incur negligible computational cost compared with the per-bit processing carried out by the syndrome-trellis decoders. To quantify the decoding complexity, we evaluate the number of Add–Compare–Select (ACS) operations required per decrypted ciphertext bit. These operations are highly amenable to efficient hardware and software implementation, enabling scalable high-throughput realizations.

Assume that the MPC cryptosystem employs a CC with memory $p$ together with a rank-$l$ masking matrix. The decryption process then requires $2^l$ parallel syndrome-trellis decoders, each operating with $2^p$ ACS units per bit. Consequently, the total ACS complexity per ciphertext bit is$2^{l+p}$. This complexity is independent of the plaintext length and therefore scales linearly with the ciphertext size, ensuring predictable performance for arbitrarily long messages.

For example, using the CC defined by (34) with $l$=6 and memory $p$=14, the total number of ACS operations per decrypted ciphertext bit is

$$2^{l+p} = 2^{20}$$

which remains well within the capabilities of modern commercial processors and dedicated hardware accelerators, including those integrated into contemporary mobile devices.

## 8. CONCLUSION

This work introduced a novel post-quantum cryptographic framework based on high-memory masked parity-check matrices of convolutional codes, addressing several inherent limitations of traditional block-code-based cryptosystems. Security is enhanced through random error injection into the plaintext, combined with dense random masking and invertible linear transformations that produce dense, random-like parity-check matrices, thereby mitigating vulnerabilities associated with low-density or structurally distinguishable codes.

A principal novelty of this work is the introduction of directed-graph decoders as the practical decryption mechanism for masked code-based crypto systems. Decoding is performed by an array of syndrome trellises, dynamically constructed by the legitimate recipient from predefined building blocks. While authorized users resolve the resulting decryption ambiguity in polynomial time, an adversary appears to be limited to exponential-time decoding attacks.

The analysis presented in this work indicates that, for the parameter sets considered, the proposed framework can achieve substantially higher estimated security margins than Niederreiter or Classic McEliece, with estimated ISD attack-complexity improvements exceeding factors of $2^{100}$against quantum adversaries and $2^{200}$against classical adversaries, while maintaining practical decoding complexity for legitimate users.

Beyond its security properties, the proposed construction supports arbitrary plaintext lengths, linear overall decoding complexity with message length, and efficient high-throughput implementation in both software and hardware. Collectively, these characteristics make the proposed MPC framework a promising candidate for robust, scalable, and practically deployable post-quantum public-key cryptography.

## AUTHOR

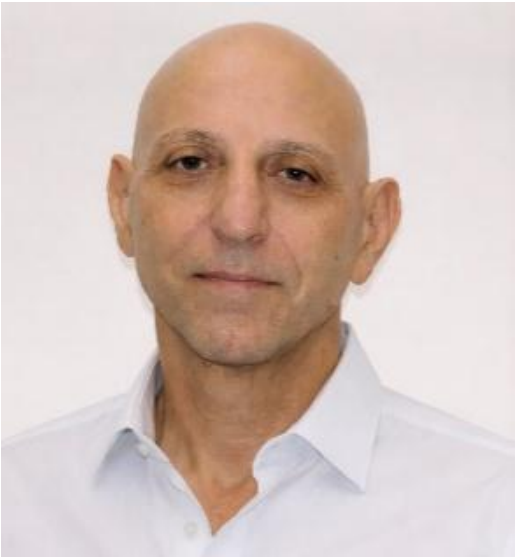

**Meir Ariel** received the B.Sc. and M.Sc. degrees (with honors) in Electrical Engineering and the Ph.D. degree in Algebraic Group Theory from Tel Aviv University, Israel. He has more than 30 years of research and development experience in signal processing, wireless communications, information theory, cryptography, and space technologies, holding leadership positions in both industry and the public sector. From 1999 to 2013, he served as Chief Executive Officer of several technology companies specializing in computer vision, Voice over IP (VoIP), and financial technologies (FinTech). Since 2013, he has been the Director General of the Herzliya Science Center. In 2018, he founded the Space Engineering Center at Tel Aviv University, where he currently serves as its Head. Under his leadership, the Center has developed and launched 24 nanosatellites for scientific research and experimental space communications. He holds 15 patents in information theory and signal processing and has served on national advisory committees in science and education. His honors include the Marco Polo Society Award (2018) and recognition by Israel's Ministry of Science and Technology as one of the country's 60 pioneering inventors (2016).